\documentclass[12pt,twoside]{article}
\usepackage[mathscr]{eucal}
\usepackage{amsmath,amsfonts,amssymb,amsthm,mathabx,empheq}
\usepackage{times}
\usepackage{pdfsync}
\usepackage{cite}
\usepackage{url}
\usepackage{hyperref}
\usepackage{tensor}
\usepackage{color}

\allowdisplaybreaks[1]

\advance\textheight\topskip
\def\nn{\nonumber}
\def\bP{\bar{P}}
\def\p{\partial}
\def\bz{\bar z}

\def\n{\nabla}
\def\cL{\mathcal{L}}
\def\cO{\mathcal{O}}

\def\half{\frac{1}{2}}

\newcommand{\bea}{\begin{eqnarray}}
\newcommand{\eea}{\end{eqnarray}}
\newcommand{\be}{\begin{equation}}
\newcommand{\ee}{\end{equation}}
\newcommand*\xbar[1]{%
  \hbox{%
    \vbox{%
      \hrule height 0.5pt 
      \kern0.3ex
      \hbox{%
        \kern-0.0em
        \ensuremath{#1}%
        \kern-0.0em
      }%
    }%
  }%
}
\def\bm{\xbar m}

\numberwithin{equation}{section} \makeatletter
\@addtoreset{equation}{section}

\DeclareFontFamily{OT1}{rsfs}{} \DeclareFontShape{OT1}{rsfs}{m}{n}{
<-7> rsfs5 <7-10> rsfs7 <10-> rsfs10}{}
\DeclareMathAlphabet{\mycal}{OT1}{rsfs}{m}{n}

\begin{document}

\title{Gravitational and electromagnetic memory}

\author{Pujian Mao and Wen-Di Tan}

\date{}

\def\mytitle{Gravitational and electromagnetic memory}

\pagestyle{myheadings} \markboth{\textsc{\small P.~Mao, W.~Tan}}
{\textsc{\small Gravitational and electromagnetic memory}}

\addtolength{\headsep}{4pt}

\begin{centering}

  \vspace{1cm}

  \textbf{\Large{\mytitle}}

  \vspace{1.5cm}

  {\large Pujian Mao and Wen-Di Tan}

\vspace{.5cm}

\vspace{.5cm}
\begin{minipage}{.9\textwidth}\small \it  \begin{center}
     Center for Joint Quantum Studies and Department of Physics,\\
     School of Science, Tianjin University, 135 Yaguan Road, Tianjin 300350, China
 \end{center}
\end{minipage}

\end{centering}


\vspace{1cm}

\begin{center}
\begin{minipage}{.9\textwidth}
  \textsc{Abstract}. We present a unified investigation of memory effect in Einstein-Maxwell theory. We specify two types of memory effect, a velocity kick and a position displacement, by examining the motion of a single free falling charged test particle. Our result recovers the two known gravitational memory effect formulas and the two known electromagnetic memory effect formulas.
 \end{minipage}
\end{center}


\thispagestyle{empty}


\section{Introduction}
\label{sec:introduction}
In the last few years, there has been renewed interest on gravitational \cite{memory,Braginsky:1986ia,1987Natur,Christodoulou:1991cr,Wiseman:1991ss,Thorne:1992sdb,Frauendiener} and electromagnetic \cite{Bieri:2013hqa} memory effects. Although both of them have been investigated for a long time  (see also \cite{Susskind:2015hpa,Lasky:2016knh,Nichols:2017rqr,Madler:2016ggp,Madler:2017umy,Madler:2018tkl,Yang:2018ceq,Bachlechner:2019deb} for the realization in experimental detections), the new enthusiasm comes from a purely theoretical side. In 2014, Strominger and Zhiboedov discovered a fundamental connection between the
gravitational memory effect and Weinberg's soft graviton theorem \cite{Strominger:2014pwa}. They are mathematically equivalent. This equivalence was shortly extended to gauge theories \cite{Pasterski:2015zua,Mao:2017wvx,Pate:2017vwa}. Inspired by this fascinating equivalence, new gravitational \cite{Pasterski:2015tva} and new electromagnetic \cite{Mao:2017axa} memory effects were reported.

The investigation in the literature on memory effect are performed independently for different theories, \textit{either gravitational memory in Einstein theory or electromagnetic memory in Maxwell theory}\footnote{Memory effect was investigated in \cite{Bieri:2010tq} in Einstein-Maxwell theory. But only gravitational effect was involved.}. A unified treatment of different types of memory effects in a coupled theory is still missing. Though gravitational memory effect and electromagnetic memory effect seem to be present at an order in which there is no coupling between the gravitational term and electromagnetic term, the main gap of connecting memory in different theories is encoded in the different types of observation. In Einstein or Maxwell theory, memory effect is interpreted as a change in the waveform of gravitational or electromagnetic wave burst. The memory effect is completely determined by the solution of Einstein equation or Maxwell's equation. The gravitational memory \cite{Frauendiener} and the new gravitational memory \cite{Pasterski:2015tva} are characterized by the change of the asymptotic shear of the outgoing null surfaces $\Delta \sigma^0$ and its  u-integral $\int \sigma^0 du$. The electromagnetic memory \cite{Bieri:2013hqa} and the new electromagnetic memory \cite{Mao:2017axa} are characterized by the change of the asymptotic data of the gauge field $\Delta A_z^0$ and its u-integral $\int A_z^0 du$. In general relativity, it is important to focus upon the coordinate invariant observable. The gravitational memory effect \cite{Strominger:2014pwa} is a relative displacement of nearby observers, while the new gravitational memory effect \cite{Pasterski:2015tva} is a relative time delay between different orbiting light rays. When we turn to the electromagnetic memory, a single charged test particle is utilized. The electromagnetic memory effect \cite{Bieri:2013hqa} is a change of the velocity (a ``kick'') of the charged particle, while the new electromagnetic memory effect \cite{Mao:2017axa} is a position displacement of the charged particle. Hence, one has to implement completely different detections to explore gravitational and electromagnetic memory effects. The aim of the present work is to provide a unified treatment for gravitational and electromagnetic memory effects in Einstein-Maxwell theory. To achieve this, we will give up the requirement of coordinate invariant observable, e.g., the proper separation between two test particles or the proper time of a single test particle. Alternatively, we will study the motion of charged particles.

Free falling observers receive a velocity kick when gravitational waves with memory pass by \cite{Zhang:2017rno,Zhang:2017jma,Zhang:2018srn,Hamada:2018cjj,Compere:2018ylh,Mao:2018xcw,Flanagan:2018yzh} (see also \cite{Grishchuk:1989qa,Podolsky:2002sa,Podolsky:2010xh,Podolsky:2016mqg} for earlier but less relevant investigations). This is the observational effect we will adopt from the gravitational side to connect with the electromagnetic memory effect. In this work, we examine the memory effect via studying the motion of a charged free falling particle\footnote{These are test particles. We do not consider them as a local source to the Einstein-Maxwell theory.}.
By solving the equations of motion, we find that the charged particle, which is initially static, is forced to orbit over some tiny angle about the ``center'' of the spacetime by the gravitational and electromagnetic radiation. The velocity change of the charged particle induced by gravitational and electromagnetic radiation is determined by $\Delta \sigma^0$ and $\Delta A_z^0$. Hence, they recover the gravitational and electromagnetic memory formulas, respectively. The position displacement of the charged particle involves u-integral of $\sigma^0$ and $A_z^0$. The gravitational and electromagnetic contributions reproduce the spin memory formula in \cite{Pasterski:2015tva} and the new electromagnetic memory formula in \cite{Mao:2017axa} respectively\footnote{The displacement effect is from a single test particle, while the displacement discovered in \cite{memory,Braginsky:1986ia,1987Natur} is a relative displacement of nearby observers. So, they are different types of memory effect.}. The charged particle receives a time delay. The contributions to the time delay are from the massive objects with or without electric charge in the spacetime \cite{Shapiro:1964uw,Visser:1998ua,Visser:1999fe}, gravitational radiation \cite{Mao:2018xcw,Flanagan:2018yzh}, and electromagnetic radiation.  The gravitational and electromagnetic memory effects happen at the same order, while the contribution of electromagnetic radiation to the time delay of the charged particle shows up at one order higher than gravitational radiation.

Our plan is as follows. In the next section, we study the Einstein-Maxwell theory in the Newman-Penrose (NP) formalism \cite{Newman:1961qr}. We work in the NP formalism because it makes the geometrical property of the spacetime more transparent. Hence, we can easily find the connection between the memory formula and the geometrical property of the spacetime. The NP formalism also has a natural connection with the spinor formalism, which is the most satisfactory way of investigating fermion coupled theories. We obtain the most general asymptotic solutions of Einstein-Maxwell theory that asymptotically approach flatness. The solution space generalizes the result of \cite{Kozarzewski,Exton:1969im} by relaxing the unit 2 sphere boundary to the case of an arbitrary 2 surface boundary, although such relaxation is not really needed for deriving the memory formulas in the present work. The solution space of Einstein-Maxwell theory allows us to derive the memory formulas and to compute the time delay of the charged particle in Section \ref{memories}. Finally, the two known gravitational memory effects and the two known electromagnetic memory effects are recovered. We then conclude with a discussion. The NP equations are listed in Appendix \ref{NPequations}.

\section{Einstein-Maxwell theory in the NP formalism}
\label{EM}

The NP formalism is a tetrad formalism where two real null vectors $e_1=l,\;e_2=n$, one complex null vector $e_3=m$ and its complex conjugate vector $e_4=\bm$ are chosen as the basis vectors. The metric is constructed from the basis vectors as
\be\label{npmetric}
g_{\mu\nu}=n_\mu l_\nu + l_\mu n_\nu - m_\mu {\bm}_\nu - m_\nu \bm_\mu.
\ee
In a hyperbolic Riemannian manifold \cite{Newman:1961qr}, it is always possible to introduce a coordinate system $(u,r,x^A)$ where $(A=z,\bz)$ and $z=e^{i\phi}\cot\frac{\theta}{2},\,\bz=e^{-i\phi}\cot\frac{\theta}{2}$ are the standard stereographic coordinates, such that the basis vectors and the cotetrad have the form
\be\label{tetrad}\begin{split}
&n^\mu \p_\mu=\frac{\p}{\p u} + U \frac{\p}{\p r} + X^A \frac{\p}{\p x^A},\;\;\;\;\;\;l^\mu \p_\mu=\frac{\p}{\p r},\;\;\;\;\;\;m^\mu \p_\mu=\omega\frac{\p}{\p r} + L^A \frac{\p}{\p x^A},\\
&n_\mu dx^\mu=\big[-U-X^A(\xbar\omega L_A+\omega \bar L_A)\big]du + dr + (\omega\bar L_A+\xbar\omega L_A)dx^A,\\
&l_\mu dx^\mu=du,\;\;\;\;\;\;m_\mu dx^\mu=-X^A L_A du + L_A dx^A,
\end{split}\ee
where $L_AL^A=0$, $L_A\bar L^A=-1$. The connection coefficients are called spin coefficients in the NP formalism with special Greek symbols (we will follow the convention of \cite{Chandrasekhar}),
\begin{align}
&\kappa=\Gamma_{311}=l^\nu m^\mu\nabla_\nu l_\mu,\;\;\pi=-\Gamma_{421}=-l^\nu \bar{m}^\mu\nabla_\nu n_\mu,\nn\\
&\epsilon=\half(\Gamma_{211}-\Gamma_{431})=\half(l^\nu n^\mu\nabla_\nu l_\mu - l^\nu \bar{m}^\mu\nabla_\nu m_\mu),\nn\\
&\tau=\Gamma_{312}=n^\nu m^\mu\nabla_\nu l_\mu,\;\;\nu=-\Gamma_{422}=-n^\nu \bar{m}^\mu\nabla_\nu n_\mu,\nn\\
&\gamma=\half(\Gamma_{212}-\Gamma_{432})=\half(n^\nu n^\mu\nabla_\nu l_\mu - n^\nu \bar{m}^\mu\nabla_\nu m_\mu),\nn\\
&\sigma=\Gamma_{313}=m^\nu m^\mu\nabla_\nu l_\mu,\;\;\mu=-\Gamma_{423}=-m^\nu \bar{m}^\mu\nabla_\nu n_\mu,\nn\\
&\beta=\half(\Gamma_{213}-\Gamma_{433})=\half(m^\nu n^\mu\nabla_\nu l_\mu - m^\nu \bar{m}^\mu\nabla_\nu m_\mu),\nn\\
&\rho=\Gamma_{314}=\bar{m}^\nu m^\mu\nabla_\nu l_\mu,\;\;\lambda=-\Gamma_{424}=-\bar{m}^\nu \bar{m}^\mu\nabla_\nu n_\mu,\nn\\
&\alpha=\half(\Gamma_{214}-\Gamma_{434})=\half(\bar{m}^\nu n^\mu\nabla_\nu l_\mu - \bar{m}^\nu \bar{m}^\mu\nabla_\nu m_\mu).\nn
\end{align}
The freedom of the rotations of the basis vectors allows one to set
\be
\pi=\kappa=\epsilon=0,\,\,\rho=\bar\rho,\,\,\tau=\bar\alpha+\beta.
\ee
Ten independent components of the Weyl tensors are represented by five complex scalars
\begin{align}
\Psi_0=-C_{1313},\;\;\Psi_1=-C_{1213},\;\;\Psi_2=-C_{1342},\;\;\Psi_3=-C_{1242},\;\;\Psi_4=-C_{2324}.\nn
\end{align}
Ricci tensors are defined in terms of four real and three complex scalars
\begin{align}
&\Phi_{00}=-\half R_{11},\;\;\Phi_{22}=-\half R_{22},\;\;\Phi_{02}=-\half R_{33},\;\;\Phi_{20}=-\half R_{44},\nn\\
&\Phi_{11}=-\dfrac{1}{4}(R_{12}+R_{34}),\;\;\Phi_{01}=-\half R_{13},\;\;,\Phi_{12}=-\half R_{23},\nn\\
&\dfrac{1}{24}R=\dfrac{1}{12} (R_{12}-R_{34}),\;\;\Phi_{10}=-\half R_{14},\;\;\Phi_{21}=-\half R_{24}.\nn
\end{align}
The Maxwell-tensor is replaced by three complex scalars
\be
\phi_0=F_{\mu\nu} l^\mu m^\nu,\quad \phi_1=\frac12 F_{\mu\nu} (l^\mu n^\nu + \bm^\mu m^\nu),\quad \phi_2=F_{\mu\nu} \bm^\mu n^\nu .\nn
\ee
The Lagrangian of four-dimensional Einstein-Maxwell theory is
\be\label{lagrangian}
\cL=\sqrt{-g}\left[ R - \frac12 F^2\right],\qquad F=dA.
\ee
For the coupled theory, $R=0$ and $\Phi_{ab}$ should be replaced by $\phi_a \xbar\phi_b$. As directional derivatives, the basis vectors are designated with special symbols
\be
D=l^\mu\p_\mu,\;\;\;\;\Delta=n^\mu\p_\mu,\;\;\;\;\delta=m^\mu\p_\mu.
\ee
The Newman-Penrose equations that we will deal with are listed in Appendix \ref{NPequations}.

The main conditions of approaching flatness at infinity are $\Psi_0=\frac{\Psi_0^0}{r^5} + \cO(r^{-6})$ and $\phi_0=\frac{\phi_0^0}{r^3} + \cO(r^{-4})$. The solutions of the NP equations in asymptotic expansions were first obtained in \cite{Kozarzewski,Exton:1969im}. However a special choice of the boundary topology $S^2$ was adopted in \cite{Exton:1969im}. We remove this restriction and a more general solution space with arbitrary 2 surface boundary topology is given by\footnote{The relaxation is encoded in the leading order of $L^{\bz}$. We have an arbitrary function $P(u,z,\bz)$ rather than a particular choice $\frac{1+z\bz}{\sqrt2}$ for a unit 2 sphere. The relaxation in the solution space is mainly shown in the integration constant \eqref{constant} and the evolution equations \eqref{evolution1}-\eqref{evolution2}.}:
\begin{align}
&\Psi_0=\frac{\Psi_0^0(u,z,\bz)}{r^5} + \frac{\Psi_0^1(u,z,\bz)}{r^6} + \cO(r^{-7}),\;\;\;\;\;\;\phi_0=\frac{\phi_0^0(u,z,\bz)}{r^3} + \frac{\phi_0^1(u,z,\bz)}{r^4} + \cO(r^{-5}),\nn\\
&\Psi_1=\frac{\Psi_1^0(u,z,\bz)}{r^4} + \frac{3\phi_0^0 \xbar \phi_1^0 - \xbar \eth \Psi_0^0}{r^5} + \cO(r^{-6}), \;\;\;\;\;\;\phi_1=\frac{\phi_1^0(u,z,\bz)}{r^2} - \frac{\xbar\eth\phi_0^0}{r^3} + \cO(r^{-4}),\nn\\
&\Psi_2=\frac{\Psi_2^0(u,z,\bz)}{r^3} + \frac{\phi_1^0 \xbar\phi_1^0 - \xbar \eth \Psi_1^0}{r^4} + \frac{1}{2r^5}\bigg[\lambda^0 \Psi_0^0  + \xbar\eth^2\Psi_0^0 + 3 \sigma^0 \xbar\sigma^0 \Psi_2^0 + 4 \Psi_1^0 \eth\xbar\sigma^0  + \xbar\sigma^0 \eth\Psi_1^0\nn\\
&\hspace{1.2cm}- 2 \phi_1^0\eth\xbar\phi_0^0 - 6\xbar\phi_1^0 \xbar\eth \phi_0^0 - 3\phi_0^0 \xbar\eth\xbar\phi_1^0 + (\gamma^0 + 3\xbar\gamma^0) \phi_0^0 \xbar\phi_0^0 + \xbar\phi_0^0\p_u\phi_0^0)\bigg] + \cO(r^{-6}),\nn\\
&\phi_2=\frac{\phi_2^0(u,z,\bz)}{r} - \frac{\xbar\eth\phi_1^0}{r^2} + \frac{\lambda^0\phi_0^0 + \sigma^0\xbar\sigma^0 \phi_2^0 + 2\phi_1^0 \eth\xbar\sigma^0 + \xbar\sigma^0\eth\phi_1^0 + \xbar\eth^2\phi_0^0}{r^3} + \cO(r^{-4})\nn\\
&\Psi_3=\frac{\Psi_3^0}{r^2} + \frac{\phi_2^0\xbar\phi_1^0 - \xbar \eth\Psi_2^0}{r^3} + \cO(r^{-4}),\;\;\;\;\;\;\Psi_4=\frac{\Psi_4^0}{r}-\frac{\xbar \eth \Psi_3^0}{r^2} + \cO(r^{-3}),\nn\\
&\rho=-\frac{1}{r}-\frac{\sigma^0\xbar\sigma^0}{r^3} + \frac{\sigma^0 \xbar\Psi_0^0 + \xbar\sigma^0 \Psi_0^0 - 6 (\sigma^0 \xbar\sigma^0)^2 - 2 \phi_0^0 \xbar\phi_0^0 }{6r^5} + \cO(r^{-6}),\nn\\
&\sigma=\frac{\sigma^0(u,z,\bz)}{r^2} + \frac{\sigma^0\sigma^0\xbar\sigma^0 - \frac12 \Psi_0^0}{r^4} - \frac{\Psi_0^1}{3r^5} + O(r^{-6}),\\
&\alpha=\frac{\alpha^0}{r}+\frac{\xbar\sigma^0\xbar\alpha^0}{r^2} + \frac{\sigma^0\xbar\sigma^0\alpha^0}{r^3}+ \frac{6\xbar\alpha^0\sigma^0(\xbar\sigma^0)^2 - \xbar\alpha^0\xbar\Psi_0^0 + \xbar\sigma^0 \Psi_1^0 - 2 \phi_1^0 \xbar\phi_0^0}{6r^4} + \cO(r^{-5}),\nn\\
&\beta=-\frac{\xbar\alpha^0}{r}-\frac{\sigma^0\alpha^0}{r^2}-\frac{\sigma^0\xbar\sigma^0\xbar\alpha^0+\half \Psi^0_1}{r^3}  + \frac{\xbar\eth\Psi_0^0 + \frac12\alpha^0\Psi_0^0 - 3\alpha^0(\sigma^0)^2\xbar\sigma^0 - 3 \phi_0^0 \xbar\phi_1^0}{3r^4}+ \cO(r^{-5}),\nn\\
&\tau=-\frac{\Psi^0_1}{2r^3} + \frac{\xbar\eth\Psi_0^0 + \frac12\sigma^0\xbar\Psi_1^0 - 4 \phi_0^0 \xbar\phi_1^0}{3r^4} + \cO(r^{-5}),\nn\\
&\mu=\frac{\mu^0}{r} - \frac{\sigma^0\lambda^0+\Psi^0_2}{r^2} + \frac{\sigma^0\xbar\sigma^0 \mu^0 + \frac12 \xbar\eth \Psi_1^0 - \phi_1^0\xbar\phi_1^0}{r^3} + \cO(r^{-4}),\nn\\
&\lambda=\frac{\lambda^0}{r} - \frac{\xbar\sigma^0 \mu^0}{r^2} + \frac{\sigma^0\xbar\sigma^0 \lambda^0 + \frac12 \xbar\sigma^0 \Psi_2^0 - \frac12 \phi_2^0 \xbar\phi_0^0}{r^3} + \cO(r^{-4}),\nn\\
&\gamma=\gamma^0-\frac{\Psi^0_2}{2r^2} + \frac{2\xbar\eth\Psi_1^0 + \alpha^0 \Psi_1^0- \xbar\alpha^0\xbar\Psi_1^0 - 6 \phi_1^0\xbar\phi_1^0} {6r^3}\nn\\
&\hspace{1.2cm} + \frac{1}{24r^4} \bigg[-3\lambda^0\Psi_0^0 - 3\xbar\eth^2\Psi_0^0 - 3\xbar\sigma^0 \eth \Psi_1^0 - 9 \sigma^0\xbar\sigma^0 \Psi_2^0 - 12 \Psi_1^0 \eth\xbar\sigma^0\nn\\
&\hspace{1.2cm}   + 4 (\xbar\alpha^0 \xbar\sigma^0 \Psi_1^0 - \alpha^0 \sigma^0 \xbar\Psi_1^0) + 2 (\xbar\alpha^0\eth\xbar\Psi_0^0 - \alpha^0\xbar\eth\Psi_0^0) + 8(\alpha^0 \phi_0^0 \xbar\phi_1^0 - \xbar\alpha^0 \xbar\phi_0^0 \phi_1^0)\nn\\
&\hspace{1.2cm} - 3 (\gamma^0 + 3 \xbar\gamma^0 ) \phi_0^0 \xbar\phi_0^0 + 12 \phi_1^0 \eth\xbar\phi_0^0 + 24 \xbar\phi_1^0 \xbar\eth\phi_0^0 + 9 \phi_0^0 \xbar\eth\xbar\phi_1^0 - 3 \xbar\phi_0^0 \p_u \phi_0^0\bigg] + \cO(r^{-5}),\nn\\
&\nu=\nu^0-\frac{\Psi^0_3}{r}+\frac{\xbar \eth \Psi^0_2 - 2 \phi_2^0 \xbar\phi_1^0}{2r^2} + \cO(r^{-3}),\nn\\
&\nn\\
&X^z=\frac{\bP\Psi_1^0}{6r^3} + \frac{\bP}{12r^4}\left(-\xbar\eth\Psi_0^0 - 2\sigma^0 \xbar\Psi_1^0 + 4 \phi_0^0 \xbar\phi_1^0\right) + \cO(r^{-5}),\nn\\
&\omega=\frac{\xbar \eth \sigma^0}{r}-\frac{\sigma^0\eth \xbar\sigma^0+\half \Psi^0_1}{r^2} + \frac{\xbar\eth\Psi_0^0 + 6 \sigma^0\xbar\sigma^0 \xbar\eth\sigma^0 + 2\sigma^0\xbar\Psi_1^0 - 4\phi_0^0 \xbar\phi_1^0}{6r^3} + \cO(r^{-4}),\nn\\
&U=-r(\gamma^0+\xbar\gamma^0) + \mu^0-\frac{\Psi^0_2 + \xbar \Psi^0_2}{2r}+\frac{\xbar\eth\Psi_1^0 + \eth\xbar\Psi_1^0 - 6\phi_1^0\xbar\phi_1^0}{6r^2} - \frac{1}{24r^3}\bigg[\lambda^0\Psi_0^0 + \xbar\lambda^0\xbar\Psi_0^0\\
&\hspace{1.2cm} + \xbar\eth^2\Psi_0^0 + \eth^2\xbar\Psi_0^0 + \xbar\sigma^0 \eth \Psi_1^0 + \sigma^0 \xbar\eth \xbar\Psi_1^0 + 3 \sigma^0\xbar\sigma^0 (\Psi_2^0 + \xbar\Psi_2^0)\nn\\
&\hspace{1.2cm}\p_u(\phi_0^0\xbar\phi_0^0) + 4(\gamma^0+\xbar\gamma^0)\phi_0^0\xbar\phi_0^0 - 12 \phi_1^0 \eth\xbar\phi_0^0 - 12 \xbar \phi_1^0 \xbar\eth\phi_0^0 - 3 \xbar\phi_0^0 \eth\phi_1^0 - 3 \phi_0^0 \xbar\eth\xbar\phi_1^0\bigg] + \cO(r^{-4}),\nn\\
&L^z=-\frac{\sigma^0 \bP(u,z,\bz)}{r^2} - \frac{\bP}{r^4}\left((\sigma^0)^2\xbar\sigma^0 - \frac16\Psi_0^0\right) + \frac{\bP \Psi_0^1}{12r^5} + \cO(r^{-6}),\nn\\
&L^{\bz}=\frac{P(u,z,\bz)}{r}+\frac{\sigma^0 \xbar\sigma^0 P}{r^3} + \frac{P}{12r^5}\left(12(\sigma^0 \xbar\sigma^0)^2 + \phi_0^0\xbar\phi_0^0 - 2\xbar\sigma^0 \Psi_0^0 - \sigma^0\xbar\Psi_0^0\right) + \cO(r^{-6}),\nn\\
&L_z=-\frac{r}{\bP} + \frac{\xbar\sigma^0 \Psi_0^0 + \phi_0^0 \xbar\phi_0^0 }{12\bP r^3} + \cO(r^{-4}),\;\;\;\;\;\; L_{\bz}=-\frac{\sigma^0}{P} + \frac{\Psi_0^0}{6Pr^2} + \frac{\Psi_0^1}{12Pr^3} + \cO(r^{-4}),\nn
\end{align}
where
\begin{align}
&\alpha^0=\half \bP \p_z \ln P,\;\;\;\;\;\;\mu^0=-\half P \bP \p_z \p_{\bz} \ln P\bP,\nn\\
&\lambda^0= \p_u{\xbar\sigma^0} + \xbar \sigma^0 (3\gamma^0 - \xbar \gamma^0),\nn\\
&\gamma^0=-\half \p_u \ln \bP,\;\;\;\;\;\;\nu^0=\xbar \eth (\gamma^0+\xbar\gamma^0),\label{constant}\\
&\Psi_2^0 - \xbar\Psi_2^0 = \xbar\eth^2\sigma^0 -\eth^2\xbar\sigma^0 + \xbar\sigma^0\xbar\lambda^0 -\sigma^0\lambda^0,\nn\\
&\Psi^0_3=\xbar \eth \mu^0 - \eth \lambda^0,\;\;\;\;\;\;\Psi^0_4=\xbar\eth \nu^0 - \p_u\lambda^0 - 4 \gamma^0 \lambda^0,\nn\\
&\nn\\
&\p_u \phi_0^0 + (\gamma^0 + 3 \xbar \gamma^0) \phi_0^0 = \eth\phi_1^0 + \sigma^0\phi_2^0,\label{evolution1}\\
&\p_u \phi_1^0 + 2(\gamma^0 + \xbar \gamma^0) \phi_1^0 = \eth\phi_2^0,\label{evolution12}\\
&\nn\\
&\p_u\Psi^0_0 + (\gamma^0 + 5 \xbar \gamma^0)\Psi^0_0=\eth\Psi^0_1 + 3\sigma^0\Psi^0_2 + 3 \phi_0^0 \xbar\phi_2^0,\\
&\p_u\Psi^0_1 + 2 (\gamma^0 + 2 \xbar \gamma^0)\Psi^0_1=\eth\Psi^0_2+2\sigma^0\Psi^0_3 + 2 \phi_1^0 \xbar\phi_2^0,\label{evolution21}\\
&\p_u\Psi^0_2 + 3 (\gamma^0 + \xbar \gamma^0)\Psi^0_2=\eth\Psi^0_3 + \sigma^0\Psi^0_4 + \phi_2^0 \xbar\phi_2^0,\label{evolution22}\\
&\p_u\Psi^0_3 + 2 (2 \gamma^0 + \xbar \gamma^0)\Psi^0_3=\eth\Psi^0_4.\label{evolution2}
\end{align}
The ``$\eth$'' operator is defined as
\begin{equation}\begin{split}
&\eth \eta^s=P\bP^{-s}\p_{\bz}(\bP^s \eta^s)=P\p_{\bz} \eta^s + 2 s\xbar\alpha^0 \eta^s,\\
&\xbar\eth \eta^s=\bP P^{s}\p_z(P^{-s} \eta^s)=\bP\p_z \eta^s -2 s \alpha^0 \eta^s,
\end{split}\end{equation}
where $s$ is the spin weight of the field $\eta$. The spin weights of relevant fields are listed in Table \ref{t1}.
\begin{table}[h]
\caption{Spin weights}\label{t1}
\begin{center}\begin{tabular}{|c|c|c|c|c|c|c|c|c|c|c|c|c|c|c|c|c|c}
\hline
& $\eth$ & $\p_u$ & $\gamma^0$ & $\nu^0$ & $\mu^0$ & $\sigma^0$ & $\lambda^0$  & $\Psi^0_4$ &  $\Psi^0_3$ & $\Psi^0_2$ & $\Psi^0_1$ & $\Psi_0^0$ &$\phi^0_2$ & $\phi^0_1$ & $\phi_0^0$   \\
\hline
s & $1$& $0$& $0$& $-1$& $0$& $2$& $-2$  &
  $-2 $&  $-1$ & $0$ & $1$ & $2$  & $-1$ & $0$ & $1$   \\
\hline
\end{tabular}\end{center}\end{table}

We will work in retarded radial gauge $A_r=0$. In terms of the gauge fields $A_\mu$, the solution of the electromagnetic fields is
\be\label{A}
A_u^0=-(\phi_1^0 + \xbar\phi_1^0) ,\;\;\;\; \p_u A_z^0 = - \frac{\phi_2^0}{\bP} ,\;\;\;\;A_z^1=-\frac{\xbar\phi_0^0}{\bP},\;\;\;\;(\p_z A_{\bz}^0 - \p_{\bz} A_z^0)=\frac{ \phi_1^0 - \xbar\phi_1^0}{P \bP },
\ee
\be
\p_u \left(\frac{A_u^0}{P\bP}\right) = \p_u (\p_z A_{\bz}^0 + \p_{\bz} A_z^0),
\ee
where
\be
A_u=\frac{A_u^0(u,z,\bz)}{r} + \cO(r^{-2}),\;\;\;\;A_z=A_z^0(u,z,\bz) + \frac{A_z^1(u,z,\bz)}{r} + \cO(r^{-2}).
\ee

\section{Memory effects}
\label{memories}
The memory effects are all encoded in the solution space derived in the previous section. To specify the observational effects, we will examine the motion of a massive charged particle. The charged particle will be constrained to a fixed radial distance $r_0$ that is very far from the gravitational and electromagnetic source, for instance constrained on the earth. The $r=r_0$ hypersurface is time-like, its induced metric can be derived easily by inserting the solution space in the previous section into \eqref{npmetric} and \eqref{tetrad}. The induced metric in series expansions is given by
\begin{multline}\label{metric}
ds^2=\left[1+\frac{\Psi_2^0 + \xbar \Psi_2^0}{r_0} - \frac{\xbar\eth\Psi_1^0 + \eth\xbar \Psi_1^0 - 6 \phi_1^0\xbar\phi_1^0}{3r_0^2} + O(r^{-3})\right]du^2 \\
- 2 \left[\frac{\eth \xbar\sigma^0}{P_s} - \frac{2 \xbar\Psi_1^0}{3P_s r_0} + O(r_0^{-2})\right] du dz
 - 2 \left[\frac{\xbar\eth \sigma^0}{P_s} - \frac{2 \Psi_1^0}{3P_s r_0} + O(r_0^{-2})\right] du d\bz \\
 - \left[2\frac{\xbar\sigma^0r}{P^2_s} - \frac{\xbar\Psi_0^0}{3P_s^2 r_0} + O(r_0^{-2})\right]dz^2- \left[2\frac{\sigma^0 r_0}{P^2_s} - \frac{\Psi_0^0}{3P_s^2 r_0} + O(r_0^{-2})\right]d\bz^2\\
 -2\left[\frac{r_0^2}{P^2_s} + \frac{\sigma^0\xbar\sigma^0}{P^2_s}  + O(r_0^{-2})\right]dzd\bz,
\end{multline}
where $P_s=\frac{1+z\bz}{\sqrt2}$. We now work in the unit 2-sphere case by setting $P=\bP=P_s$. The induced Maxwell field on the $r=r_0$ hypersurface is
\be\begin{split}
F_{uz}=-\frac{\phi_2^0}{P_s}  + \frac{\xbar\eth\phi_1^0 - \xbar\sigma^0\xbar\phi_2^0}{P_s r_0}& + \cO(r_0^{-2}),\;\;\;\;F_{u\bz}=-\frac{\xbar\phi_2^0}{P_s}  + \frac{\eth\xbar\phi_1^0 - \sigma^0 \phi_2^0}{P_s r_0}+ \cO(r_0^{-2}),\\
F_{z\bz}=&\frac{\phi_1^0 - \xbar\phi_1^0}{P_s^2}  + \frac{\eth\xbar\phi_0^0 - \xbar\eth \phi_0^0}{P_s^2 r_0} + \cO(r_0^{-2}).
\end{split}\ee

A free falling charged particle with a net charge $q$ on this hypersurface will of course not travel along the geodesic. The tangent vector $V$ of the particle worldline satisfies
\be\label{geodesic}
V^\nu(\xbar\n_\nu V^\mu + q {\xbar F_\nu}^\mu) =0,
\ee
where $\xbar\n$ is the covariant derivative on this three-dimensional hypersurface. Following \cite{Mao:2018xcw}, we impose that $V$ is given in series expansion as
\be\label{expansion}
V^u=1 + \sum\limits_{a=1}^\infty\frac{V^u_a}{r^a},\;\;\;\;V^z=\sum\limits_{a=2}^\infty\frac{V^z_a}{r^{a}}.
\ee
Then, we can solve \eqref{geodesic} order by order. The solution up to relevant order is
\begin{align}
&V^u_1=-\frac{\Psi_2^0 + \xbar \Psi_2^0}{2},\\
&V^u_2=\frac16(\xbar\eth\Psi_1^0 + \eth \xbar\Psi_1^0) - \eth\xbar\sigma^0\xbar\eth\sigma^0 + \frac38(\Psi_2^0 + \xbar \Psi_2^0)^2 -\phi_1^0\xbar\phi_1^0 + q^2 P_s^2 A_z^0 A_{\bz}^0,\\
&V^z_2=-P_s \xbar\eth\sigma^0 + q P_s^2 A_{\bz}^0 ,\\
&V^z_3=P_s\left[2\eth\xbar\sigma^0\sigma^0 + \frac23\Psi_1^0 + \frac12\xbar\eth\sigma^0 (\Psi_2^0 + \xbar \Psi_2^0) \right]-P_s\int\;dv\; \frac{\eth(\Psi_2^0 + \xbar \Psi_2^0 + 2 q A_u^0)}{2}\nn \\
&\hspace{7.5cm}-2q P^2_s \sigma^0 A_z^0 + q P^2_s A_z^1.
\end{align}
We have set all integration constants of $u$ to zero as we require that the charged particle is initially static.

At $r_0^{-2}$ order, $V$ has angular components due to the presence of gravitational waves characterized by $\sigma^0$ and electromagnetic waves characterized by $A_z^0$. In other words, the radiation forces the charged particle to rotate over some tiny angle about the ``center'' of the spacetime $r=0$. The memory effect is the velocity kick of the charged particle
\be\label{leading}
\Delta V^z=-\frac{1}{r_0^2}(P_s \xbar\eth\Delta\sigma^0 - q P_s^2 \Delta A_{\bz}^0) + \cO(r_0^{-3}) .
\ee
It includes two parts: namely, the gravitational contribution $-P_s \xbar\eth\Delta\sigma^0$ and electromagnetic contribution $ q P_s^2 \Delta A_{\bz}^0$. They precisely recover the gravitational memory formula in \cite{Frauendiener} and the electromagnetic memory formula in \cite{Bieri:2013hqa}.

Both gravitational and electromagnetic radiation have a decomposition into the E-mode and B-mode \cite{Winicour:2014ska}.
The decomposition into electric and magnetic parts is achieved by relating $\sigma^0$ or $\phi_2^0$ to spin-weight-0 fields
\be
\sigma^0=\eth^2 \left[A(u,z,\bz) + i B(u,z,\bz)\right],\quad \phi_2^0= \p_u \xbar \eth \left[C(u,z,\bz) + i D(u,z,\bz)\right],\nn
\ee
where the second relation is equivalent to
\be
A_z^0 = - \p_z (C + i D).\nn
\ee
Inserting those decomposition into \eqref{evolution12} and \eqref{evolution22}, one obtains
\be\label{emode}
\begin{split}
&\eth\xbar\eth \Delta C = \frac12 \Delta(\phi_1^0 + \xbar \phi_1^0),\\
&\eth^2\xbar\eth^2 \Delta A= - \frac12\Delta (\Psi_2^0 + \xbar \Psi_2^0 + \sigma^0 \p_u \xbar\sigma^0 + \xbar\sigma^0 \p_u \sigma^0) + \int du(\p_u\sigma^0\p_u\xbar\sigma^0 + \phi_2^0\xbar\phi_2^0),
\end{split}
\ee
and
\be\label{bmode}
\begin{split}
&i \eth\xbar\eth \Delta D = \frac12 \Delta(\phi_1^0 - \xbar \phi_1^0),\\
&i \eth^2\xbar\eth^2 \Delta B= \frac12\Delta (\Psi_2^0 - \xbar \Psi_2^0 + \sigma^0 \p_u \xbar\sigma^0 - \xbar\sigma^0 \p_u \sigma^0) .
\end{split}
\ee
Note that we now work in the unit 2-sphere case. The E-mode electromagnetic memory in \eqref{emode} only has the ordinary part $\frac12 \Delta(\phi_1^0 + \xbar \phi_1^0)$ following the classification of \cite{Bieri:2013hqa}, because there is no charged matter coupled to the theory. Hence no charged radiation reaches null infinity. The E-mode gravitational memory has both \cite{Bieri:2013ada,Satishchandran:2019pyc} the ordinary part
\be
-\frac12\Delta (\Psi_2^0 + \xbar \Psi_2^0 + \sigma^0 \p_u \xbar\sigma^0 + \xbar\sigma^0 \p_u \sigma^0),\nn
\ee
and the null part
\be
\int du(\p_u\sigma^0\p_u\xbar\sigma^0 + \phi_2^0\xbar\phi_2^0).\nn
\ee
The B-mode memory \eqref{bmode} can not be studied from a purely asymptotic argument \cite{Winicour:2014ska}. The B-mode memory just recovers the relations
\be\begin{split}
&(\p_z A_{\bz}^0 - \p_{\bz} A_z^0)=\frac{ \phi_1^0 - \xbar\phi_1^0}{P \bP },\nn\\
&\xbar\eth^2\sigma^0 -\eth^2\xbar\sigma^0 =\Psi_2^0 - \xbar\Psi_2^0  + \sigma^0\lambda^0 - \xbar\sigma^0\xbar\lambda^0
\end{split}\ee
in \eqref{A} and \eqref{constant}. However, the B-mode memory can been seen in the position displacement as we will show below.

Following the treatment in electromagnetism \cite{Mao:2017axa}, one can define a second memory effect by a position displacement of the charged particle
\be
\Delta z=\int V^z du=-\frac{1}{r_0^2}\int du (P_s \xbar\eth \sigma^0 - q P_s^2 A_{\bz}^0) + \cO(r_0^{-3}),
\ee
where we have used the fact that $du=d\chi + \cO(r_0^{-1})$, and $\chi$ is the proper time. It also includes two parts, namely the gravitational contribution $-\int (P_s \xbar\eth \sigma^0) du$ and electromagnetic contribution $\int ( q P_s^2 A_{\bz}^0) du$. They precisely recover the spin memory formula in \cite{Pasterski:2015tva} and the displacement memory formula in \cite{Mao:2017axa}.

Inserting the B-mode decomposition of electromagnetic and gravitational radiation into \eqref{evolution1} and \eqref{evolution21}, we obtain
\be\label{bmode2}
i \xbar\eth\eth\xbar\eth\eth \int D  du = \frac12 \Delta (\xbar\eth \phi_0^0 - \eth \xbar\phi_0^0) + \frac12 \int du \left[ \eth (\xbar\sigma^0 \xbar\phi_0^0) - \xbar\eth(\sigma^0 \phi_2^0)\right],
\ee
and
\begin{multline}
i\xbar\eth \eth \xbar\eth^2\eth^2 \int B du = \frac12 \Delta (\xbar\eth\Psi_1^0 - \eth \xbar\Psi_1^0) + \frac12 \int du \left(\sigma^0 \p_u \xbar\sigma^0 - \xbar\sigma^0 \p_u \sigma^0\right)\\
+ \int du\left[\sigma^0\xbar\eth\eth \xbar\sigma^0 - \xbar\sigma^0 \eth\xbar\eth \sigma^0 + \eth (\xbar\phi_1^0 \phi_2^0) - \xbar\eth (\phi_1^0\xbar\phi_2^0) \right] .
\end{multline}
Interestingly, the B-mode electromagnetic memory \eqref{bmode2} now has a null part. The mixed term $\sigma^0 \phi_2^0$ in \eqref{evolution1} is the ``magnetic'' source that reaches null infinity.

Another observational memory effect is a time delay of the free falling particle \cite{Mao:2018xcw,Flanagan:2018yzh}. The electromagnetic radiation can also contribute to the time delay of a charged particle. Since $V$ is time-like, the infinitesimal change of the proper time can be derived from the co-vector\footnote{We have used the fact that $dz=\frac{V^z_0}{r_0^2}du + \cO(r_0^{-3})$.}
\begin{multline}
d\chi=\bigg[1  + \frac{1}{2r_0} (\Psi^0_2+\xbar\Psi^0_2 ) - \frac{1}{r_0^2}\bigg(\frac18(\Psi^0_2 + \xbar\Psi^0_2)^2 + \frac16 (\xbar\eth\Psi^0_1 + \eth\xbar\Psi^0_1) -  \xbar\eth\sigma^0 \eth\xbar\sigma^0\\
  -  \phi_1^0\xbar\phi_1^0 +  q^2 P^2_s A_z^0 A_{\bz}^0 \bigg) \bigg] du  + \cO(r_0^{-3}).
\end{multline}
Clearly, the electromagnetic contribution $ (\phi_1^0\xbar\phi_1^0 - q^2 P^2_s A_z^0 A_{\bz}^0)$ comes one order higher than the gravitational contribution $\frac12(\Psi^0_2 + \xbar \Psi^0_2)$ in the $\frac{1}{r_0}$ expansion.

\section{Discussion}
\label{sec:discussion}

In this work, the gravitational memory effect and the electromagnetic memory effect are investigated in a unified fashion by examining the motion of a charged test particle. Some interesting applications and open questions may cross the reader's mind. We have only concerned ourselves with the memory effects that are related to soft theorems in the present work. However, as reported in \cite{Compere:2019odm}, the memory effect can be defined as infinite towers at every order. We believe that the unified method we proposed here is also applicable for the higher-order memory effect. One just needs to check more orders in \eqref{expansion}. Since our motivation is to provide a unified treatment of memory effect in coupled theories. It would be of interest to test our treatment in more generic theories with more matter fields coupled in various ways or even string theory \cite{Afshar:2018sbq}. And the equivalence between soft theorems and memory effects could be investigated in a systematical way with our treatment. In the present work, we applied the Newman-Unti gauge \cite{Newman:1962cia} which is the most convenient one to derive the solution space and hence the memory effect. However, the universality of the leading soft theorems implies a gauge independent deviation of the memory effect, e.g., symmetry or conformal structure \cite{Ashtekar:2014zsa}. It is of interest to study this issue elsewhere.
Another interesting point is about the double soft theorem (see, e.g. \cite{Klose:2015xoa,Volovich:2015yoa}). Hopefully, our treatment can shine light on the understanding of the memory effect which is connected to the double soft theorem.

\section*{Acknowledgements}
\label{sec:acknowledgements}

\addcontentsline{toc}{section}{Acknowledgments}

The authors would like to thank the anonymous referees for the suggestions and comments which were very helpful in improving the original
manuscript. This work is supported in part by the NSFC (National Natural Science Foundation of China) under Grants No. 11905156 and No. 11935009.

\appendix

\section{NP equations}
\label{NPequations}

\textbf{Radial equations}
\bea
&&D\rho =\rho^2+\sigma\xbar\sigma + \phi_0 \xbar\phi_0,\label{R1}\\
&&D\sigma=2\rho \sigma + \Psi_{0},\label{R2}\\
&&D\tau =\tau \rho +  \xbar \tau \sigma   + \Psi_1 + \phi_0 \xbar\phi_1,\label{R3}\\
&&D\alpha=\rho  \alpha + \beta \xbar \sigma  + \phi_1 \xbar\phi_0,\label{R4}\\
&&D\beta  =\alpha \sigma + \rho  \beta + \Psi_{1},\label{R5}\\
&&D\gamma=\tau \alpha +  \xbar \tau \beta  + \Psi_2 + \phi_1 \xbar\phi_1,\label{R6}\\
&&D\lambda=\rho\lambda + \xbar\sigma\mu + \phi_2 \xbar\phi_0,\label{R7}\\
&&D\mu =\rho \mu + \sigma\lambda + \Psi_{2} ,\label{R8}\\
&&D\nu =\xbar\tau \mu + \tau  \lambda + \Psi_3 + \phi_2 \xbar\phi_1,\label{R9}\\
&&DU=\xbar\tau\omega+\tau\xbar\omega - (\gamma+\xbar\gamma),\label{R10}\\
&&DX^A=\xbar\tau L^A + \tau\bar L^A,\label{R11}\\
&&D\omega=\rho\omega+\sigma\xbar\omega-\tau,\label{R12}\\
&&DL^A=\rho L^A + \sigma \bar L^A,\label{R13}\\
&&D\Psi_1 - \xbar\delta \Psi_0 =  4 \rho \Psi_1 - 4\alpha \Psi_0 + \xbar \phi_1 D \phi_0 - \xbar\phi_0 \delta\phi_0 - 2\sigma \phi_1 \xbar\phi_0 + 2 \beta \phi_0 \xbar\phi_0,\label{R14}\\
&&D\Psi_2 - \xbar\delta \Psi_1 =   3\rho \Psi_2  - 2 \alpha \Psi_1- \lambda \Psi_0\nn\\
&&\hspace{1.8cm} + \xbar \phi_1 \xbar\delta \phi_0 - \xbar\phi_0 \Delta\phi_0 - 2\alpha \phi_0 \xbar\phi_1 + 2 \rho \phi_1 \xbar\phi_1 + 2 \gamma \phi_0 \xbar\phi_0 - 2 \tau\phi_1 \xbar\phi_0,\label{R15}\\
&&D\Psi_3 - \xbar\delta \Psi_2 =  2\rho \Psi_3 - 2\lambda \Psi_1 + \xbar \phi_1 D \phi_2 - \xbar\phi_0 \delta\phi_2 + 2\mu \phi_1 \xbar\phi_0 - 2 \beta \phi_2 \xbar\phi_0,\label{R16}\\
&&D\Psi_4 - \xbar\delta \Psi_3 = \rho  \Psi_4 + 2 \alpha \Psi_3 - 3 \lambda \Psi_2\nn\\
&&\hspace{1.5cm} - \xbar \phi_0 \Delta \phi_2 + \xbar\phi_1 \xbar\delta\phi_2 + 2\alpha \phi_2 \xbar\phi_1 + 2 \nu \phi_1 \xbar\phi_0 - 2 \gamma \phi_2 \xbar\phi_0 - 2 \lambda\phi_1 \xbar\phi_1,\label{R17}\\
&&D\phi_1 - \xbar\delta \phi_0 = 2\rho \phi_1 - 2\alpha \phi_0,\label{R18}\\
&&D\phi_2 - \xbar\delta \phi_1 = \rho \phi_2 - \lambda \phi_0.\label{R19}
\eea

\textbf{Non-radial  equations}
\bea
&&\Delta\lambda  = \xbar\delta\nu- (\mu + \xbar\mu)\lambda - (3\gamma - \xbar\gamma)\lambda + 2\alpha \nu - \Psi_4,\label{H1}\\
&&\Delta\rho= \xbar\delta\tau- \rho\xbar\mu - \sigma\lambda  -2\alpha \tau + (\gamma + \xbar\gamma)\rho  - \Psi_2 ,\label{H2}\\
&&\Delta\alpha = \xbar\delta\gamma +\rho \nu - (\tau + \beta)\lambda + (\xbar\gamma - \gamma -\xbar \mu)\alpha  -\Psi_3 ,\label{H3}\\
&&\Delta \mu=\delta\nu-\mu^2 - \lambda\xbar\lambda - (\gamma + \xbar\gamma)\mu   + 2 \beta \nu - \phi_2 \xbar\phi_2,\label{H4}\\
&&\Delta \beta=\delta\gamma - \mu\tau + \sigma\nu + \beta(\gamma - \xbar\gamma -\mu) - \alpha\xbar\lambda - \phi_1 \xbar\phi_2,\label{H5}\\
&&\Delta \sigma=\delta\tau - \sigma\mu - \rho\xbar\lambda - 2 \beta \tau + (3\gamma - \xbar\gamma)\sigma  - \phi_0 \xbar\phi_2,\label{H6}\\
&&\Delta \omega=\delta U +\xbar\nu -\xbar\lambda\xbar\omega + (\gamma-\xbar\gamma-\mu)\omega,\label{H7}\\
&&\Delta L^A=\delta X^A - \xbar\lambda \bar L^A + (\gamma-\xbar\gamma-\mu)L^A,\label{H8}\\
&&\delta\rho - \xbar\delta\sigma=\rho\tau - \sigma (3\alpha - \xbar\beta)   - \Psi_1 + \phi_0 \xbar\phi_1,\label{H9}\\
&&\delta\alpha - \xbar\delta\beta=\mu\rho - \lambda\sigma + \alpha\xbar\alpha + \beta\xbar\beta - 2 \alpha\beta - \Psi_2  + \phi_1 \xbar\phi_1,\label{H10}\\
&&\delta\lambda - \xbar\delta\mu= \mu \xbar\tau + \lambda (\xbar\alpha - 3\beta) - \Psi_3 + \phi_2 \xbar\phi_1,\label{H11}\\
&&\delta \xbar\omega-\bar\delta\omega=\mu - \xbar\mu - (\alpha - \xbar\beta) \omega +  (\xbar\alpha - \beta)\xbar\omega,\label{H12}\\
&&\delta \bar L^A - \bar\delta L^A= (\xbar\alpha - \beta)\bar L^A -  (\alpha - \xbar\beta) L^A ,\label{H13}\\
&&\Delta\Psi_0 - \delta \Psi_1 = (4\gamma -\mu)\Psi_0 - (4\tau + 2\beta)\Psi_1 + 3\sigma \Psi_2\nn\\
&&\hspace{5cm}- \xbar \phi_2 D \phi_0 + \xbar\phi_1 \delta\phi_0 - 2\beta \phi_0 \xbar\phi_1 + 2 \sigma \phi_1 \xbar\phi_1,\label{H14}\\
&&\Delta\Psi_1 - \delta \Psi_2 = \nu\Psi_0 + (2\gamma - 2\mu)\Psi_1 - 3\tau \Psi_2 + 2\sigma \Psi_3 \nn\\
&&\hspace{1.5cm} + \xbar \phi_1 \Delta \phi_0 - \xbar\phi_2 \xbar\delta\phi_0 - 2\rho \phi_1 \xbar\phi_2 - 2 \gamma \phi_0 \xbar\phi_1 + 2 \tau \phi_1 \xbar\phi_1 + 2 \alpha \phi_0 \xbar\phi_2,\label{H15}\\
&&\Delta\Psi_2 - \delta \Psi_3 = 2\nu \Psi_1 - 3\mu \Psi_2 + (2\beta - 2\tau) \Psi_3 + \sigma \Psi_4\nn\\
&&\hspace{5cm}- \xbar \phi_2 D \phi_2 + \xbar\phi_1 \delta\phi_2 - 2\mu \phi_1 \xbar\phi_1 + 2 \beta \phi_2 \xbar\phi_1,\label{H16}\\
&&\Delta\Psi_3 - \delta \Psi_4 = 3\nu \Psi_2 - (2\gamma + 4\mu) \Psi_3 + (4\beta - \tau) \Psi_4 \nn\\
&&\hspace{1.5cm}+ \xbar \phi_1 \Delta \phi_2 - \xbar\phi_2 \xbar\delta\phi_2 - 2\alpha \phi_2 \xbar\phi_2 - 2 \nu \phi_1 \xbar\phi_1 + 2 \gamma \phi_2 \xbar\phi_1 + 2 \lambda\phi_1 \xbar\phi_2,\label{H17}\\
&&\Delta\phi_0 - \delta \phi_1 = (2\gamma-\mu) \phi_0 - 2\tau \phi_1 + \sigma \phi_2,\label{H18}\\
&&\Delta\phi_1 - \delta \phi_2 = \nu \phi_0 - 2 \mu \phi_1 - (\xbar\alpha - \beta) \phi_2.\label{H19}
\eea


\begin{thebibliography}{10}

\bibitem{memory}
Y.~B. Zel'dovich and A.~G. Polnarev, ``{Radiation of gravitational waves by a
  cluster of superdense stars},'' {\em Soviet Astronomy} {\bfseries 18} (Aug.,
  1974) 17.

\bibitem{Braginsky:1986ia}
V.~B. Braginsky and L.~P. Grishchuk, ``{Kinematic Resonance and Memory Effect
  in Free Mass Gravitational Antennas},'' {\em Sov. Phys. JETP} {\bfseries 62}
  (1985) 427--430.
[Zh. Eksp. Teor. Fiz.89,744(1985)].

\bibitem{1987Natur}
V.~B. {Braginskii} and K.~S. {Thorne}, ``{Gravitational-wave bursts with memory
  and experimental prospects},'' \href{http://dx.doi.org/10.1038/327123a0}{{\em
  Nature} {\bfseries 327} (May, 1987) 123--125}.

\bibitem{Christodoulou:1991cr}
D.~Christodoulou, ``{Nonlinear nature of gravitation and gravitational wave
  experiments},''
\href{http://dx.doi.org/10.1103/PhysRevLett.67.1486}{{\em Phys. Rev. Lett.}
  {\bfseries 67} (1991) 1486--1489}.

\bibitem{Wiseman:1991ss}
A.~G. Wiseman and C.~M. Will, ``{Christodoulou's nonlinear gravitational wave
  memory: Evaluation in the quadrupole approximation},''
\href{http://dx.doi.org/10.1103/PhysRevD.44.R2945}{{\em Phys. Rev.} {\bfseries
  D44} no.~10, (1991) R2945--R2949}.

\bibitem{Thorne:1992sdb}
K.~S. Thorne, ``{Gravitational-wave bursts with memory: The Christodoulou
  effect},''
\href{http://dx.doi.org/10.1103/PhysRevD.45.520}{{\em Phys. Rev.} {\bfseries
  D45} no.~2, (1992) 520--524}.

\bibitem{Frauendiener}
J.~Frauendiener, ``{Note on the memory effect},''
  \href{http://dx.doi.org/10.1088/0264-9381/9/6/018}{{\em Class. Quant. Grav.}
  {\bfseries 9} (1992) 1639--1641}.

\bibitem{Bieri:2013hqa}
L.~Bieri and D.~Garfinkle, ``{An electromagnetic analogue of gravitational wave
  memory},'' \href{http://dx.doi.org/10.1088/0264-9381/30/19/195009}{{\em
  Class. Quant. Grav.} {\bfseries 30} (2013) 195009},
\href{http://arxiv.org/abs/1307.5098}{{\ttfamily arXiv:1307.5098 [gr-qc]}}.

\bibitem{Susskind:2015hpa}
L.~Susskind, ``{Electromagnetic Memory},''
\href{http://arxiv.org/abs/1507.02584}{{\ttfamily arXiv:1507.02584 [hep-th]}}.

\bibitem{Lasky:2016knh}
P.~D. Lasky, E.~Thrane, Y.~Levin, J.~Blackman, and Y.~Chen, ``{Detecting
  gravitational-wave memory with LIGO: implications of GW150914},''
  \href{http://dx.doi.org/10.1103/PhysRevLett.117.061102}{{\em Phys. Rev.
  Lett.} {\bfseries 117} no.~6, (2016) 061102},
\href{http://arxiv.org/abs/1605.01415}{{\ttfamily arXiv:1605.01415
  [astro-ph.HE]}}.

\bibitem{Nichols:2017rqr}
D.~A. Nichols, ``{Spin memory effect for compact binaries in the post-Newtonian
  approximation},'' \href{http://dx.doi.org/10.1103/PhysRevD.95.084048}{{\em
  Phys. Rev.} {\bfseries D95} no.~8, (2017) 084048},
\href{http://arxiv.org/abs/1702.03300}{{\ttfamily arXiv:1702.03300 [gr-qc]}}.

\bibitem{Madler:2016ggp}
T.~M\"{a}dler and J.~Winicour, ``{The sky pattern of the linearized gravitational
  memory effect},''
  \href{http://dx.doi.org/10.1088/0264-9381/33/17/175006}{{\em Class. Quant.
  Grav.} {\bfseries 33} no.~17, (2016) 175006},
\href{http://arxiv.org/abs/1605.01273}{{\ttfamily arXiv:1605.01273 [gr-qc]}}.

\bibitem{Madler:2017umy}
T.~M\"{a}dler and J.~Winicour, ``{Radiation Memory, Boosted Schwarzschild
  Spacetimes and Supertranslations},''
  \href{http://dx.doi.org/10.1088/1361-6382/aa6ca8}{{\em Class. Quant. Grav.}
  {\bfseries 34} no.~11, (2017) 115009},
\href{http://arxiv.org/abs/1701.02556}{{\ttfamily arXiv:1701.02556 [gr-qc]}}.

\bibitem{Madler:2018tkl}
T.~M\"{a}dler and J.~Winicour, ``{Kerr Black Holes and Nonlinear Radiation
  Memory},'' \href{http://dx.doi.org/10.1088/1361-6382/ab1187}{{\em Class.
  Quant. Grav.} {\bfseries 36} no.~9, (2019) 095009},
\href{http://arxiv.org/abs/1811.04711}{{\ttfamily arXiv:1811.04711 [gr-qc]}}.

\bibitem{Yang:2018ceq}
H.~Yang and D.~Martynov, ``{Testing Gravitational Memory Generation with
  Compact Binary Mergers},''
  \href{http://dx.doi.org/10.1103/PhysRevLett.121.071102}{{\em Phys. Rev.
  Lett.} {\bfseries 121} no.~7, (2018) 071102},
  \href{http://arxiv.org/abs/1803.02429}{{\ttfamily arXiv:1803.02429 [gr-qc]}}.

\bibitem{Bachlechner:2019deb}
T.~C. Bachlechner and M.~Kleban, ``{Testing the electric Aharonov-Bohm effect
  with superconductors},''
  \href{http://dx.doi.org/10.1103/PhysRevB.101.174504}{{\em Phys. Rev. B}
  {\bfseries 101} no.~17, (2020) 174504},
  \href{http://arxiv.org/abs/1909.11668}{{\ttfamily arXiv:1909.11668
  [hep-th]}}.

\bibitem{Strominger:2014pwa}
A.~Strominger and A.~Zhiboedov, ``{Gravitational Memory, BMS Supertranslations
  and Soft Theorems},'' \href{http://dx.doi.org/10.1007/JHEP01(2016)086}{{\em
  JHEP} {\bfseries 01} (2016) 086},
\href{http://arxiv.org/abs/1411.5745}{{\ttfamily arXiv:1411.5745 [hep-th]}}.

\bibitem{Pasterski:2015zua}
S.~Pasterski, ``{Asymptotic Symmetries and Electromagnetic Memory},''
  \href{http://dx.doi.org/10.1007/JHEP09(2017)154}{{\em JHEP} {\bfseries 09}
  (2017) 154},
\href{http://arxiv.org/abs/1505.00716}{{\ttfamily arXiv:1505.00716 [hep-th]}}.

\bibitem{Mao:2017wvx}
P.~Mao and H.~Ouyang, ``{Note on soft theorems and memories in even
  dimensions},'' \href{http://dx.doi.org/10.1016/j.physletb.2017.08.064}{{\em
  Phys. Lett.} {\bfseries B774} (2017) 715--722},
\href{http://arxiv.org/abs/1707.07118}{{\ttfamily arXiv:1707.07118 [hep-th]}}.

\bibitem{Pate:2017vwa}
M.~Pate, A.-M. Raclariu, and A.~Strominger, ``{Color Memory: A Yang-Mills
  Analog of Gravitational Wave Memory},''
  \href{http://dx.doi.org/10.1103/PhysRevLett.119.261602}{{\em Phys. Rev.
  Lett.} {\bfseries 119} no.~26, (2017) 261602},
\href{http://arxiv.org/abs/1707.08016}{{\ttfamily arXiv:1707.08016 [hep-th]}}.

\bibitem{Pasterski:2015tva}
S.~Pasterski, A.~Strominger, and A.~Zhiboedov, ``{New Gravitational
  Memories},'' \href{http://dx.doi.org/10.1007/JHEP12(2016)053}{{\em JHEP}
  {\bfseries 12} (2016) 053},
\href{http://arxiv.org/abs/1502.06120}{{\ttfamily arXiv:1502.06120 [hep-th]}}.

\bibitem{Mao:2017axa}
P.~Mao, H.~Ouyang, J.-B. Wu, and X.~Wu, ``{New electromagnetic memories and
  soft photon theorems},''
  \href{http://dx.doi.org/10.1103/PhysRevD.95.125011}{{\em Phys. Rev.}
  {\bfseries D95} no.~12, (2017) 125011},
\href{http://arxiv.org/abs/1703.06588}{{\ttfamily arXiv:1703.06588 [hep-th]}}.

\bibitem{Bieri:2010tq}
L.~Bieri, P.~Chen, and S.-T. Yau, ``{Null asymptotics of solutions of the
  Einstein-Maxwell equations in general relativity and gravitational
  Radiation},'' \href{http://dx.doi.org/10.4310/ATMP.2011.v15.n4.a5}{{\em Adv.
  Theor. Math. Phys.} {\bfseries 15} no.~4, (2011) 1085--1113},
\href{http://arxiv.org/abs/1011.2267}{{\ttfamily arXiv:1011.2267 [math.DG]}}.

\bibitem{Zhang:2017rno}
P.~M. Zhang, C.~Duval, G.~W. Gibbons, and P.~A. Horvathy, ``{The Memory Effect
  for Plane Gravitational Waves},''
  \href{http://dx.doi.org/10.1016/j.physletb.2017.07.050}{{\em Phys. Lett.}
  {\bfseries B772} (2017) 743--746},
\href{http://arxiv.org/abs/1704.05997}{{\ttfamily arXiv:1704.05997 [gr-qc]}}.

\bibitem{Zhang:2017jma}
P.~M. Zhang, C.~Duval, and P.~A. Horvathy, ``{Memory Effect for Impulsive
  Gravitational Waves},''
  \href{http://dx.doi.org/10.1088/1361-6382/aaa987}{{\em Class. Quant. Grav.}
  {\bfseries 35} no.~6, (2018) 065011},
\href{http://arxiv.org/abs/1709.02299}{{\ttfamily arXiv:1709.02299 [gr-qc]}}.

\bibitem{Zhang:2018srn}
P.~M. Zhang, C.~Duval, G.~W. Gibbons, and P.~A. Horvathy, ``{Velocity Memory
  Effect for Polarized Gravitational Waves},''
  \href{http://dx.doi.org/10.1088/1475-7516/2018/05/030}{{\em JCAP} {\bfseries
  1805} no.~05, (2018) 030},
\href{http://arxiv.org/abs/1802.09061}{{\ttfamily arXiv:1802.09061 [gr-qc]}}.

\bibitem{Hamada:2018cjj}
Y.~Hamada and S.~Sugishita, ``{Notes on the gravitational, electromagnetic and
  axion memory effects},''
  \href{http://dx.doi.org/10.1007/JHEP07(2018)017}{{\em JHEP} {\bfseries 07}
  (2018) 017},
\href{http://arxiv.org/abs/1803.00738}{{\ttfamily arXiv:1803.00738 [hep-th]}}.

\bibitem{Compere:2018ylh}
G.~Comp\`{e}re, A.~Fiorucci, and R.~Ruzziconi, ``{Superboost transitions,
  refraction memory and super-Lorentz charge algebra},''
  \href{http://dx.doi.org/10.1007/JHEP11(2018)200}{{\em JHEP} {\bfseries 11}
  (2018) 200},
\href{http://arxiv.org/abs/1810.00377}{{\ttfamily arXiv:1810.00377 [hep-th]}}.

\bibitem{Mao:2018xcw}
P.~Mao and X.~Wu, ``{More on gravitational memory},''
  \href{http://dx.doi.org/10.1007/JHEP05(2019)058}{{\em JHEP} {\bfseries 05}
  (2019) 058},
\href{http://arxiv.org/abs/1812.07168}{{\ttfamily arXiv:1812.07168 [gr-qc]}}.

\bibitem{Flanagan:2018yzh}
E.~E. Flanagan, A.~M. Grant, A.~I. Harte, and D.~A. Nichols, ``{Persistent
  gravitational wave observables: general framework},''
  \href{http://dx.doi.org/10.1103/PhysRevD.99.084044}{{\em Phys. Rev.}
  {\bfseries D99} no.~8, (2019) 084044},
\href{http://arxiv.org/abs/1901.00021}{{\ttfamily arXiv:1901.00021 [gr-qc]}}.

\bibitem{Grishchuk:1989qa}
L.~P. Grishchuk and A.~G. Polnarev, ``{Gravitational wave pulses with 'velocity
  coded memory.'},'' {\em Sov. Phys. JETP} {\bfseries 69} (1989) 653--657.
[Zh. Eksp. Teor. Fiz.96,1153(1989)].

\bibitem{Podolsky:2002sa}
J.~Podolsky and R.~Steinbauer, ``{Geodesics in space-times with expanding
  impulsive gravitational waves},''
  \href{http://dx.doi.org/10.1103/PhysRevD.67.064013}{{\em Phys. Rev.}
  {\bfseries D67} (2003) 064013},
\href{http://arxiv.org/abs/gr-qc/0210007}{{\ttfamily arXiv:gr-qc/0210007
  [gr-qc]}}.

\bibitem{Podolsky:2010xh}
J.~Podolsky and R.~Svarc, ``{Refraction of geodesics by impulsive spherical
  gravitational waves in constant-curvature spacetimes with a cosmological
  constant},'' \href{http://dx.doi.org/10.1103/PhysRevD.81.124035}{{\em Phys.
  Rev.} {\bfseries D81} (2010) 124035},
\href{http://arxiv.org/abs/1005.4613}{{\ttfamily arXiv:1005.4613 [gr-qc]}}.

\bibitem{Podolsky:2016mqg}
J.~Podolsky, C.~Samann, R.~Steinbauer, and R.~Svarc, ``{The global uniqueness
  and $C^1$-regularity of geodesics in expanding impulsive gravitational
  waves},'' \href{http://dx.doi.org/10.1088/0264-9381/33/19/195010}{{\em Class.
  Quant. Grav.} {\bfseries 33} no.~19, (2016) 195010},
\href{http://arxiv.org/abs/1602.05020}{{\ttfamily arXiv:1602.05020 [gr-qc]}}.

\bibitem{Shapiro:1964uw}
I.~I. Shapiro, ``{Fourth Test of General Relativity},''
\href{http://dx.doi.org/10.1103/PhysRevLett.13.789}{{\em Phys. Rev. Lett.}
  {\bfseries 13} (1964) 789--791}.

\bibitem{Visser:1998ua}
M.~Visser, B.~Bassett, and S.~Liberati, ``{Superluminal censorship},''
  \href{http://dx.doi.org/10.1016/S0920-5632(00)00782-9}{{\em Nucl. Phys. Proc.
  Suppl.} {\bfseries 88} (2000) 267--270},
\href{http://arxiv.org/abs/gr-qc/9810026}{{\ttfamily arXiv:gr-qc/9810026
  [gr-qc]}}.

\bibitem{Visser:1999fe}
M.~Visser, B.~Bassett, and S.~Liberati, ``{Perturbative superluminal censorship
  and the null energy condition},''
  \href{http://dx.doi.org/10.1063/1.1301601}{{\em AIP Conf. Proc.} {\bfseries
  493} no.~1, (1999) 301--305},
\href{http://arxiv.org/abs/gr-qc/9908023}{{\ttfamily arXiv:gr-qc/9908023
  [gr-qc]}}.

\bibitem{Newman:1961qr}
E.~Newman and R.~Penrose, ``{An Approach to gravitational radiation by a method
  of spin coefficients},''
\href{http://dx.doi.org/10.1063/1.1724257}{{\em J. Math. Phys.} {\bfseries 3}
  (1962) 566--578}.

\bibitem{Kozarzewski}
B.~Kozarzewski, ``{Asymptotic properties of the electromagnetic and
  gravitational fields},'' {\em Acta Phys. Polon.} {\bfseries 27} (1965) 775.

\bibitem{Exton:1969im}
A.~R. Exton, E.~T. Newman, and R.~Penrose, ``{Conserved quantities in the
  Einstein-Maxwell theory},''
\href{http://dx.doi.org/10.1063/1.1665006}{{\em J. Math. Phys.} {\bfseries 10}
  (1969) 1566--1570}.

\bibitem{Chandrasekhar}
S.~Chandrasekhar, ``{The Newman-Penrose formalism},'' in {\em {The mathematical
  theory of black holes}}, ch.~1, pp.~40--55.
\newblock Oxford, UK, 1983.

\bibitem{Winicour:2014ska}
J.~Winicour, ``{Global aspects of radiation memory},''
  \href{http://dx.doi.org/10.1088/0264-9381/31/20/205003}{{\em Class. Quant.
  Grav.} {\bfseries 31} (2014) 205003},
\href{http://arxiv.org/abs/1407.0259}{{\ttfamily arXiv:1407.0259 [gr-qc]}}.

\bibitem{Bieri:2013ada}
L.~Bieri and D.~Garfinkle, ``{Perturbative and gauge invariant treatment of
  gravitational wave memory},''
  \href{http://dx.doi.org/10.1103/PhysRevD.89.084039}{{\em Phys. Rev. D}
  {\bfseries 89} no.~8, (2014) 084039},
  \href{http://arxiv.org/abs/1312.6871}{{\ttfamily arXiv:1312.6871 [gr-qc]}}.

\bibitem{Satishchandran:2019pyc}
G.~Satishchandran and R.~M. Wald, ``{Asymptotic behavior of massless fields and
  the memory effect},''
  \href{http://dx.doi.org/10.1103/PhysRevD.99.084007}{{\em Phys. Rev. D}
  {\bfseries 99} no.~8, (2019) 084007},
  \href{http://arxiv.org/abs/1901.05942}{{\ttfamily arXiv:1901.05942 [gr-qc]}}.

\bibitem{Compere:2019odm}
G.~Comp\`{e}re, ``{Infinite towers of supertranslation and superrotation
  memories},'' \href{http://dx.doi.org/10.1103/PhysRevLett.123.021101}{{\em
  Phys. Rev. Lett.} {\bfseries 123} no.~2, (2019) 021101},
\href{http://arxiv.org/abs/1904.00280}{{\ttfamily arXiv:1904.00280 [gr-qc]}}.

\bibitem{Afshar:2018sbq}
H.~Afshar, E.~Esmaeili, and M.~Sheikh-Jabbari, ``{String Memory Effect},''
  \href{http://dx.doi.org/10.1007/JHEP02(2019)053}{{\em JHEP} {\bfseries 02}
  (2019) 053}, \href{http://arxiv.org/abs/1811.07368}{{\ttfamily
  arXiv:1811.07368 [hep-th]}}.

\bibitem{Newman:1962cia}
E.~T. Newman and T.~W.~J. Unti, ``{Behavior of Asymptotically Flat Empty
  Spaces},''
\href{http://dx.doi.org/10.1063/1.1724303}{{\em J. Math. Phys.} {\bfseries 3}
  no.~5, (1962) 891}.

\bibitem{Ashtekar:2014zsa}
A.~Ashtekar, ``{Geometry and Physics of Null Infinity},''
  \href{http://arxiv.org/abs/1409.1800}{{\ttfamily arXiv:1409.1800 [gr-qc]}}.

\bibitem{Klose:2015xoa}
T.~Klose, T.~McLoughlin, D.~Nandan, J.~Plefka, and G.~Travaglini,
  ``{Double-Soft Limits of Gluons and Gravitons},''
  \href{http://dx.doi.org/10.1007/JHEP07(2015)135}{{\em JHEP} {\bfseries 07}
  (2015) 135}, \href{http://arxiv.org/abs/1504.05558}{{\ttfamily
  arXiv:1504.05558 [hep-th]}}.

\bibitem{Volovich:2015yoa}
A.~Volovich, C.~Wen, and M.~Zlotnikov, ``{Double Soft Theorems in Gauge and
  String Theories},'' \href{http://dx.doi.org/10.1007/JHEP07(2015)095}{{\em
  JHEP} {\bfseries 07} (2015) 095},
\href{http://arxiv.org/abs/1504.05559}{{\ttfamily arXiv:1504.05559 [hep-th]}}.

\end{thebibliography}

\providecommand{\href}[2]{#2}\begingroup\raggedright\endgroup

\end{document}